**Title**: Ecosystem service demand relationship and trade-off patterns in urban parks across China

**Authors**: Shuyao Wu[1#], Delong Li[2#], Ximan Sun[1], Kai-di Liu[3], Wentao Zhang[4], Binbin V. Li[5,6], Shuangcheng Li[7,8], Lumeng Liu[1], Fangjin Xu[7,8], Jinwei Dong[2], Laibao Liu[9], Weili Duan[10], Zhonghao Zhang[1*]

[1] College of Geography and Remote Sensing, Hohai University, Nanjing, Jiangsu, 211100, China
[2] Institute of Geographic Sciences and Natural Resources Research, Chinese Academy of Science, Beijing, 100101, China
[3] School of Innovation and Entrepreneurship, Shandong University, Qingdao, Shandong 266237, China
[4] Institute of Humanities and Arts, Shandong University, Qingdao, Shandong 266237, China
[5] Environmental Research Centre, Duke Kunshan University, Kunshan, Jiangsu 215316, China. [6] Nicholas School of the Environment, Duke University, Durham, NC 27708, USA
[7] College of Urban and Environmental Sciences, Peking University, Beijing 100871, China
[8] Key Laboratory for Earth Surface Processes of the Ministry of Education, Peking University, Beijing 100871, China
[9] Department of Geography, Institute for Climate and Carbon Neutrality, The University of Hong Kong, Hong Kong, 999077, China
[10] Xinjiang Institute of Ecology and Geography, Chinese Academy of Science, Urumqi, Xinjiang, 830011, China

[#] These authors contributed equally.

*** Corresponding to**:
zzh87@hhu.edu.cn (Zhonghao Zhang)

**Email addresses**:
wushuyao@hhu.edu.cn (Shuyao Wu)
lidelong@igsnrr.ac.cn (Delong Li)
250218010018@hhu.edu.cn (Ximan Sun)
liukaidi@sdu.edu.cn (Kai-di Liu)
zhangwt@sdu.edu.cn (Wentao Zhang)
binbin.li@duke.edu (Binbin V. Li)
scli@urban.pku.edu.cn (Shuangcheng Li)
liulumeng@hhu.edu.cn (Lumeng Liu)
xu_fj@stu.pku.edu.cn (Fangjin Xu)
dongjw@igsnrr.ac.cn (Jinwei Dong)
laibao@hku.hk (Laibao Liu)
duanweili@ms.xjb.ac.cn (Weili Duan)
zzh87@hhu.edu.cn (Zhonghao Zhang)

**Acknowledgments**
We thank the National Natural Science Foundation of China (42501366, 42301344) for funding this research.

**Data availability**
All data used for identifying influential factors are publicly available, as described in Supplementary Table S1. Processed survey data on ecosystem service demand is available at https://doi.org/10.5281/zenodo.17309624. The analytical code supporting the findings of this study is deposited at https://github.com/lidelong8/ES_demand.

**Conflict of interests**
The authors declare that they have no known competing financial interests or personal relationships that could have appeared to influence the work reported in this paper.

**Abstract**

Understanding public demand for urban ecosystem services (ES) is crucial for effective green space management, yet the intricate relationships and potential trade-offs among these diverse demands remain poorly understood. Previous studies have yielded inconsistent findings, often limited by small samples or reliance on indirect proxies. Here, we provide the first national-scale, direct assessment of the relationship among demands for nine urban park ES using a survey dataset comprising 20,075 responses across China and a point-allotment experiment that directly quantifies the trade-off patterns among service demands. We found particularly strong preferences among urban residents in China for air purification and recreation services, at the expense of other services. These preferences were further reflected in three distinct demand bundles: air purification-dominated, recreation-dominated, and balanced demands, each delineating a typical group of people with distinct representative characteristics. Socio-economic and environmental factors, such as age, environmental interest, and mean annual precipitation, significantly influence the trade-off intensity among service demands. Our study pioneers the direct, quantitative analysis of relationships among ecosystem service demands, and the results underscore the need for tailored urban park designs that address diverse service demands to sustainably enhance the quality of city life in China and beyond.

## 1. Introduction

Ecosystem services (ES), which refer to the benefits people obtain from nature, are essential for the existence and quality of people's lives (Brauman et al., 2020; Costanza et al., 2017). In cities across the globe, urban green infrastructures, such as urban parks and green spaces, also provide city dwellers with various important ecosystem services, including air pollution reduction, local temperature and humidity regulation, noise abatement, biodiversity conservation, etc. (Evans et al., 2022; Bakhtsiyarava et al., 2024; Rocha et al., 2024; Ma and Yang, 2025). These ecosystem services derive from neutral ecosystem functions, turn into services that are valuable to people, and ultimately improve human well-being (Haines-Young and Potschin, 2010). The key to turning those neutral ecosystem functions into services is the demand for the services from people (Haines-Young and Potschin, 2010). Without the demands, ecosystem functions would remain natural processes and be less, at least directly, impactful on human societies. Therefore, understanding people's demands for various ecosystem services is vital for understanding the nature of human-nature interactions. Without a thorough understanding of ecosystem service demand (ESD), assessments of ES supply lack context and societal relevance, making it difficult to identify potential resource conflicts, ensure equitable distribution of benefits, or design effective management interventions (De Knegt et al., 2025; Gao et al., 2026).

Mounting evidence has shown that the *supply* of ecosystem services exhibits complex relationships of synergies or trade-offs, often forming distinct "bundles" (Wu and Li, 2019; Hua et al., 2022; Wang et al., 2024). For example, at the global scale, Wang et al. (2024) showed that strong synergies exist among the supply of services such as microclimate regulation and carbon sequestration, as well as many provisioning and cultural services, while significant trade-offs are found between services like flood regulation and water conservation or soil retention. However, whether the *demands* for ecosystem services also present such relationships or patterns remains largely unknown. Unlike supply, which is rooted in biophysical processes, demand is a socio-economic construct that is inherently complex, multi-faceted, and dynamic (He et al., 2025). While a few studies have attempted to address the problem using survey results, their findings have been inconsistent, often constrained by limited survey samples or by reliance on proxies to represent demand levels for various ecosystem services (Fang et al., 2024). For instance, Washbourne et al. (2020) identified strong trade-offs between the five types of ecosystem service demands based on a survey of 140 respondents in the UK, whereas Yuan et al. (2023) uncovered predominantly synergistic relationships in China using a range of indices to represent the demand for four ecosystem services (i.e., population density for food production, water consumption per unit of GDP for water conservation, actual amount of soil erosion, and nighttime light for carbon sequestration). More recently, Huang et al. (2026) used land-use types as a proxy for ESD and showed that conflicts may arise among societal demands for food production, water conservation, carbon sequestration, and habitat protection. These contradictory outcomes not only hamper our understanding of the relationships among ES demands but also undermine the development of effective environmental planning that can enhance the well-being of the massive urban population worldwide (Luederitz et al., 2015; Cortinovis and Geneletti, 2019).

Here, we address this important gap by employing a direct assessment method to measure the demand for various ecosystem services in China (Wu et al., 2026b). China serves as a highly suitable study area for understanding the relationships among urban ecosystem service demands (Li et al., 2020, 2024). Its rapid urbanization has created unprecedented demand for urban ES, while its top-down planning approach means that comprehensive knowledge of service demand can directly inform green space management (Jia, 2023). Therefore, we applied the dataset from a national-scale survey experiment in China spanning a diverse range of urban environments to capture the nuanced interrelationships and trade-offs among nine ecosystem service demands in urban parks. By using a point allotment experiment with over 20,000 responses, we aim to answer the following three key research questions:

(1) How do the demands for different ecosystem services in urban parks correlate with one another across China?

(2) What constitutes the typical bundles of ecosystem service demands, and what are the demographic profiles of the individuals associated with these bundles?

(3) What drivers underpin the trade-off intensities among the various ecosystem service demands?

## 2. Methods

### 2.1 Ecosystem service demand data

Demand data for ecosystem services were obtained from a national online questionnaire survey on ecosystem service preferences across China from July 2022 to January 2024, previously published in Wu et al., 2026a and 2026b. The questionnaire aimed to assess people's preferences for nine ecosystem services (i.e., Air Purification, Local Climate Regulation, Noise Attenuation, Flood Mitigation, Recreation, Education, Food and Water Supply, Habitat Maintenance, Others) commonly found in urban parks using a point-allotment experiment design. In the point-allotment experiment, respondents were instructed to assign a total of 100 importance points across the nine ES to reflect their perceived importance of the ES, with individual service points ranging from 0 to 100 (i.e., higher allotted points indicate stronger service demand). This method explicitly quantifies both the ordinal ranking and the intensity of service preferences by imposing a maximum number of points (Khan et al., 2022; Patrycja et al., 2022).

The variations among allotted service points, as indicated by the measure of standard deviation, were used to represent the trade-off intensities among ESD (i.e., a higher standard deviation value indicates a less even value distribution and implies stronger trade-off intensities among services) (Martín-López et al., 2012). Moreover, we used Spearman rank correlation analysis to examine the relationship between ESD (i.e., the received importance points for each service) and the K-means clustering technique, identifying potential service demand bundles based on service importance points. The silhouette method and gap statistic indicate that the ideal number of clusters is approximately three (Mouchet et al., 2014; Mur et al., 2016; Luo et al., 2024).

Information on respondent socio-demographic characteristics (i.e., gender, age, income level, education level, and city of residence) and personal preferences (i.e., self-reported interest in various environmental issues, monthly frequency of park visits, and the size of the most frequently visited urban park) was also collected in the survey. After strict quality screening (filtering invalid responses by completion time, logical consistency, and attention checks), 20,075 valid responses were retained, covering 31 provincial regions and 344 prefecture-level cities in mainland China (Fig. S1). A supplementary set of 2,233 responses with unusually short or long completion times was used for sensitivity testing. More detailed descriptions of the dataset can be found in the related work of Wu et al., 2026b, and it is available at https://doi.org/10.5281/zenodo.17309624.

## 2.2 Influential factors identification

We selected 13 variables to identify potential key factors influencing ESD trade-off intensities (Table S1). These included five socio-demographic and personal variables (gender, age, education, income, environmental interest), two park preference variables (monthly frequency of park visits and the size of the most frequently visited urban park), four local environmental variables (mean annual temperature, mean annual precipitation, mean annual PM2.5 concentration, and urban vegetation coverage), and two city-level socio-economic variables (population density and GDP density). These factors are found to influence the demand for ecosystem services in urban parks both theoretically and empirically (Bakhtsiyarava et al., 2024; Wu et al., 2026a). The detailed rationale and data sources for each factor are provided in Table S1.

To assess the importance of these factors on trade-off intensities, we employed the eXtreme Gradient Boosting (XGBoost) algorithm. The XGBoost is an ensemble learning method based on decision trees that builds models sequentially, with each new model correcting the errors of its predecessor (Parsa et al., 2020). It is widely recognized for its high predictive accuracy and robustness in handling complex, non-linear relationships within diverse datasets (Zhang et al., 2020; Xia et al., 2026). To be specific, the key hyperparameters for the XGBoost regressor were set as follows: nrounds = 200 (total number of decision trees), eta = 0.1 (learning rate), max_depth = 6 (maximum depth of a single tree), subsample = 0.8 (fraction of samples used per tree), colsample_bytree = 0.8 (fraction of features used per tree), min_child_weight = 1, alpha = 0 (L1 regularization), lambda = 1 (L2 regularization), and seed = 0 (random seed for reproducibility). Furthermore, we used Shapley Additive exPlanations (SHAP) values to interpret the XGBoost model's results (Ekanayake et al., 2022; Xia et al., 2026). The SHAP values, derived from game theory, provide a unified measure of feature importance by quantifying each feature's contribution to an individual prediction (Park et al., 2022; Zhou et al., 2022). This approach allows for both global interpretation, by ranking features based on their overall impact, and local interpretation, by explaining specific predictions (Park et al., 2022; Zhou et al., 2022). In addition, to examine how each influential factor affected the trade-off intensities, we generated accumulated local effects (ALE) plots (Apley and Zhu, 2020; Seihani et al., 2024). The ALE plots show the average marginal effect of a feature on the model's prediction, are more robust to potential feature correlations, and thus provide a more accurate and unbiased representation of the feature's influence (Apley and Zhu, 2020; Seihani et al., 2024). To better visualize and interpret the trends in the ALE plots, we also fitted the commonly used nonparametric Locally Estimated Scatterplot Smoothing (LOESS) curve to each plot (Xue et al., 2025). All analyses were implemented in R software (version 4.2.2) using the xgboost and iml packages.

## 3. Results

### 3.1 Ecosystem service demand relationship

Across all 20,075 responses, the strongest trade-off relationships were observed between demands for Air Purification and Education (Spearman's $\rho = -0.38$) and between Air Purification and Food and Water Supply ($\rho = -0.38$). The strongest synergy was between Education and Food and Water Supply ($\rho = 0.19$) (Fig. 1a and S3). The overall trade-off intensity (mean standard deviation) was 9.15. We identified three typical demand bundles: air purification-dominated (19.31% of respondents), recreation-dominated (22.00%), and balanced demands (58.69%). Within these bundles, the strongest trade-offs were found between Noise Attenuation and Recreation (-0.36) in the air purification-dominated bundle, Noise Attenuation and Habitat Maintenance (-0.30) in the recreation-dominated bundle, and Noise Attenuation and Local Climate Regulation (-0.42) in the balanced bundle (Fig. 1b to 1d, S3). The strongest synergies were between Noise Attenuation and Local Climate Regulation in the air purification-dominated (0.06) and recreation-dominated (0.12) bundles, and between Education and Food and Water Supply in the balanced bundle (0.16) (Fig. 1b to 1d, S3). Overall, the correlation coefficients were moderate, with the strongest trade-off and synergy being -0.42 and 0.19, respectively. The sensitivity analysis results suggested that the relationships among ecosystem service demands were robust, which showed no significant differences in the main conclusions (Fig. S4).

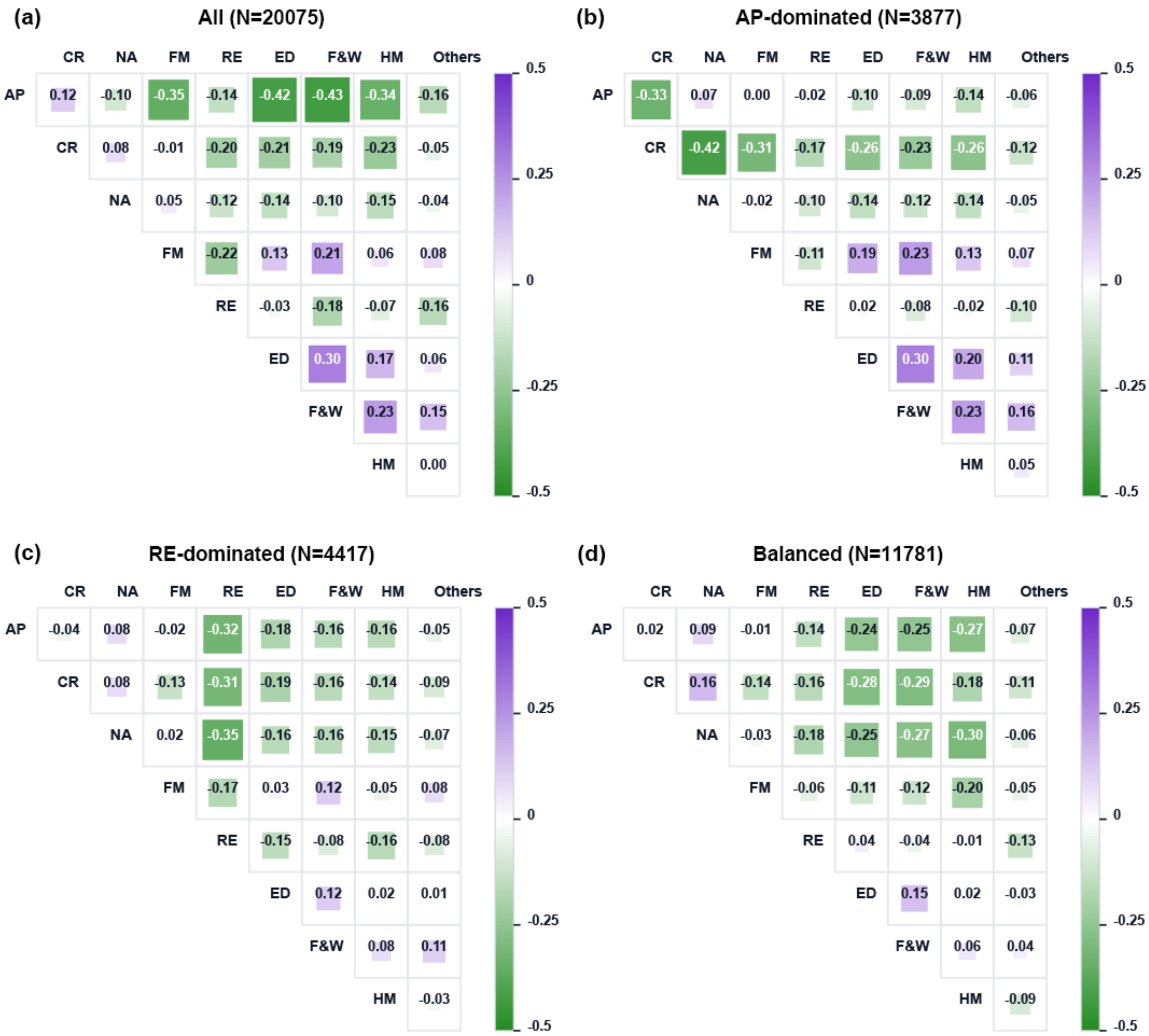

Fig. 1. Spearman rank correlation matrices between nine ecosystem service demands in urban parks within 20,075 respondents (a) and three typical demand bundles (b to d). AP stands for air purification; CR stands for local climate regulation; NA stands for noise attenuation; FM stands for flood mitigation; RE stands for recreation; ED stands for education; F&W stands for food and water supply; HM stands for habitat maintenance. Color and size of squares correspond to the strength and direction of correlation, respectively.

Among all services, the demand for Air Purification received the highest average importance points (22.52) (Fig. 2a and S2). On the other hand, the demand for Food and Water Supply received the lowest importance points (6.55), except for Others (2.44 on average, only 22 valid responses, including ESD of soil retention, wind shelter, biological control, and spiritual experiences). The mean importance points of the other six services in descending order (i.e., Recreation, Local Climate Regulation, Noise Attenuation, Habitat Maintenance, Education, and Flood Mitigation) were 15.68, 13.55, 12.49, 9.59, 8.54, and 8.46, respectively (Fig. 2a and S2). In the air purification-dominated and recreation-dominated bundles, the Air Purification and Recreation services received 42.38 and 28.71 importance points on average, respectively, which were much higher than the means of the other services in the same bundles (7.20 and 8.91, respectively) (Fig. 2a and S2). The balanced bundle exhibited a more even distribution of service demands, with the highest importance point (16.73) for Air Purification and the lowest (8.77) for Food and Water Supply, which were closer to the bundle's mean importance point (11.08). The service demand trade-off intensities were also significantly higher in the air purification-dominated and recreation-dominated bundles than in the balanced bundle and all responses pooled (Fig. 2b).

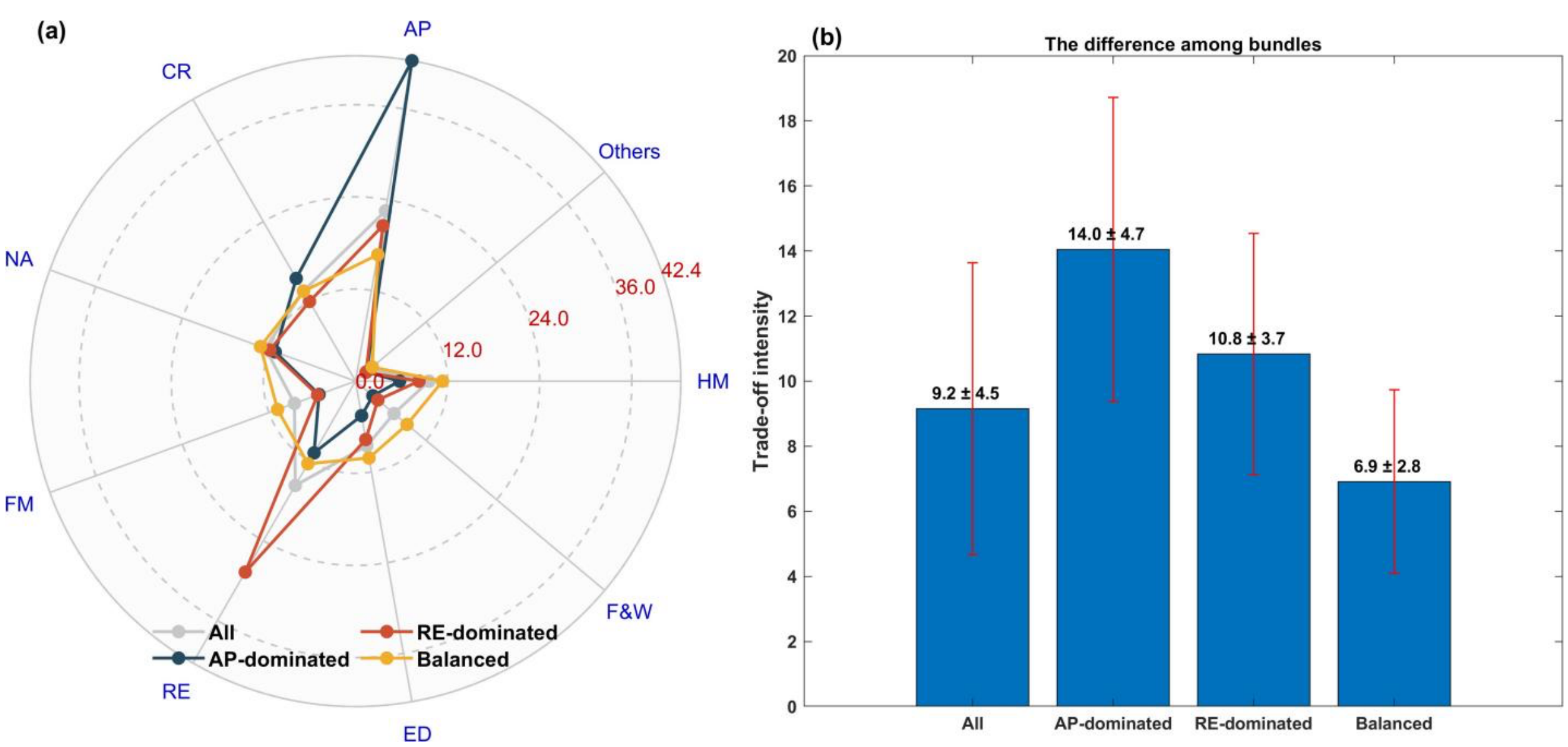

Fig. 2. Average ecosystem service importance points allotted by all 20,075 respondents and within three typical ecosystem service demand bundles (a) and the average trade-off intensity (i.e., standard deviation values) (b). AP stands for air purification; CR stands for local climate regulation; NA stands for noise attenuation; FM stands for flood mitigation; RE stands for recreation; ED stands for education; F&W stands for food and water supply; HM stands for habitat maintenance. All differences in trade-off intensities were significant ($p < 0.001$) based on Kolmogorov-Smirnov tests. Error bars represent ±1 standard error.

### 3.2 Service demand bundle profile

To understand the differences in the typical profile of people within each bundle, we showed the distribution frequency of the 13 variables among respondents and applied the Kolmogorov-Smirnov tests to determine whether the distributions of variables differ significantly among bundles (Fig. 3). For the gender difference, we found significantly more women in the balanced bundle than in the air purification-dominated bundle (men were coded one and women were coded two in the analysis; thus, the higher average value suggested more women in the bundle, Fig. 3a). For the age difference, older people were more found in the recreation-dominated bundle while more younger people were found in the balanced bundle (Fig. 3b). For the differences in education and income levels, there were more people with higher education and income levels in the recreation-dominated bundle (people with higher income levels were also equally common in the air-purification dominated bundle) (Fig. 3c and 3d). On average, the balanced bundle also contained a higher number of individuals who reported stronger interests in environmental issues (Fig. 3e). In terms of the differences in environmental variables, people who live in cities with higher vegetation coverage were more found in the balanced bundle (Fig. 3i). On the other hand, people from more populous cities with lower $PM_{2.5}$ concentrations and higher GDP levels were more common in the recreation-dominated bundle (Fig. 3h, 3j, and 3k).

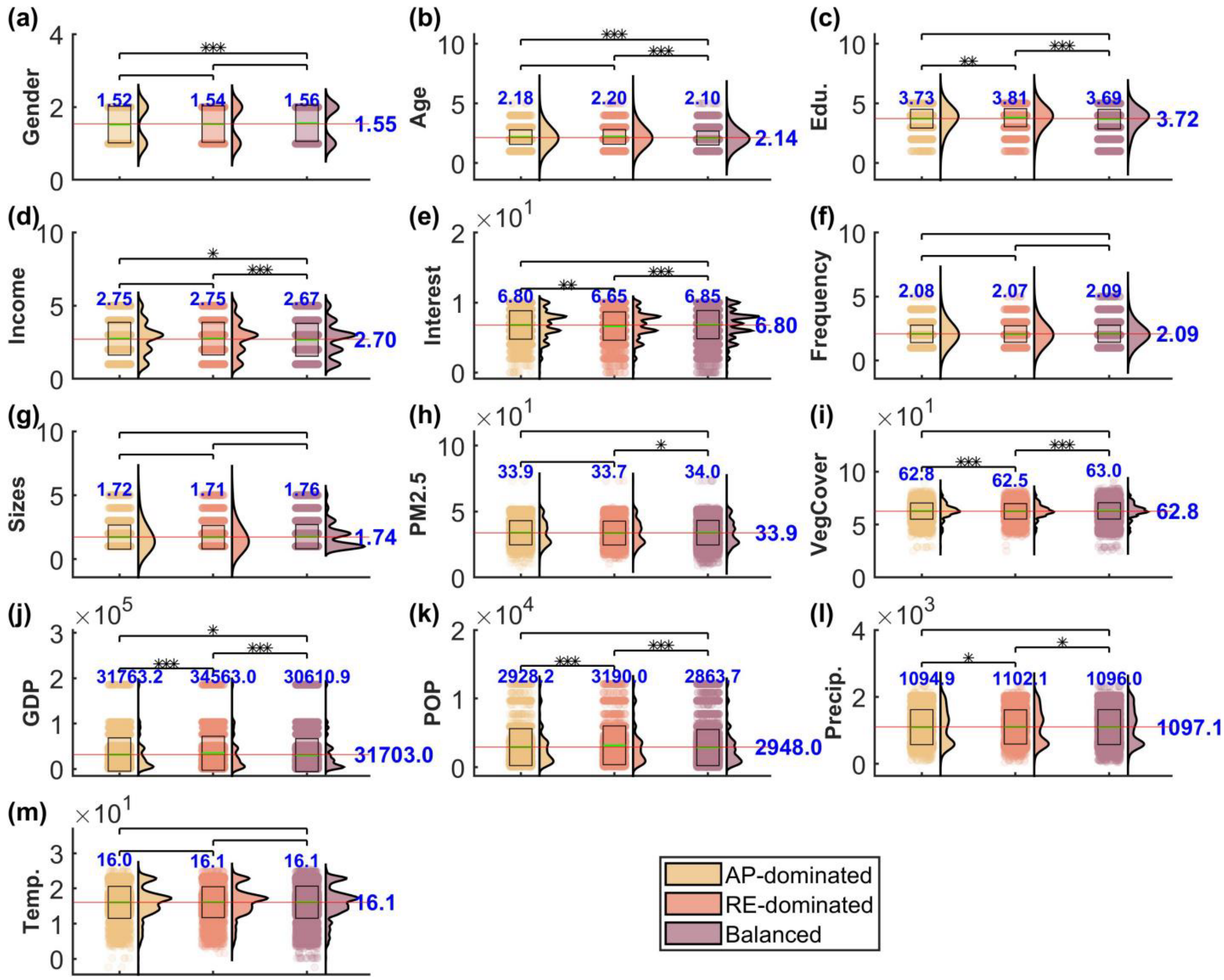

Fig. 3. Distribution frequency of the 13 socio-economic, personal, and environmental variables in three ecosystem service demand bundles. Lines and values indicate the means of the variables in each bundle (green lines) and all respondents (red lines). Interest represents respondents' interest in environmental issues; Sizes represent the size of the most frequently visited park; Frequency is the monthly frequency of park visits; Edu. stands for education level; VegCover stands for vegetation coverage; POP stands for population density; Temp. stands for mean annual temperature; Precip. stands for mean annual precipitation. *** $p < 0.001$, ** $p < 0.01$, * $p < 0.05$ based on Kolmogorov-Smirnov test results.

## 3.3 Influential factors of demand trade-off intensity

All selected factors passed the collinearity test, with Variance Inflation Factor (VIF) values below 10, confirming their suitability for inclusion in the model. The XGBoost-SHAP analysis revealed that a combination of personal attributes, local environmental conditions, and city-level socioeconomic background collectively shapes how individuals perceive and prioritize competing ecosystem services in urban parks. The most important influential factors of trade-off intensity were identified as people's age, interest in environmental issues, and park visit frequency for socio-economic or personal variables and mean annual precipitation, population density, and mean annual temperature for environmental variables (Fig. 4a). Income level and vegetation coverage appeared to be the least influential personal and environmental factors, respectively (Fig. 4a). Accumulated local effects analyses further showed nonlinear marginal responses of trade-off intensity to 13 environmental, socio-economic, and personal variables (Fig. 4b–n). Older people who reported lower interest in environmental issues tended to prefer fewer services (i.e., higher trade-off intensity). In comparison, younger people who were strongly interested in various environmental issues had more balanced preferences for different services (i.e., lower trade-off intensity) (Fig. 4b and 4c). Men who prefer to visit larger parks but rarely do so appeared to have higher ESD trade-offs, suggesting that their preferences focus on a smaller number of services (Fig. 4d, 4e, and 4f). Additionally, people who live in cities with mean annual precipitation between 500 and 1500 mm per year, medium-sized population density (between 1,500 and 5,000 people per $km^2$), and higher mean annual temperature were found to have lower trade-off intensities among service demands (Fig. 4g, 4h, and 4j).

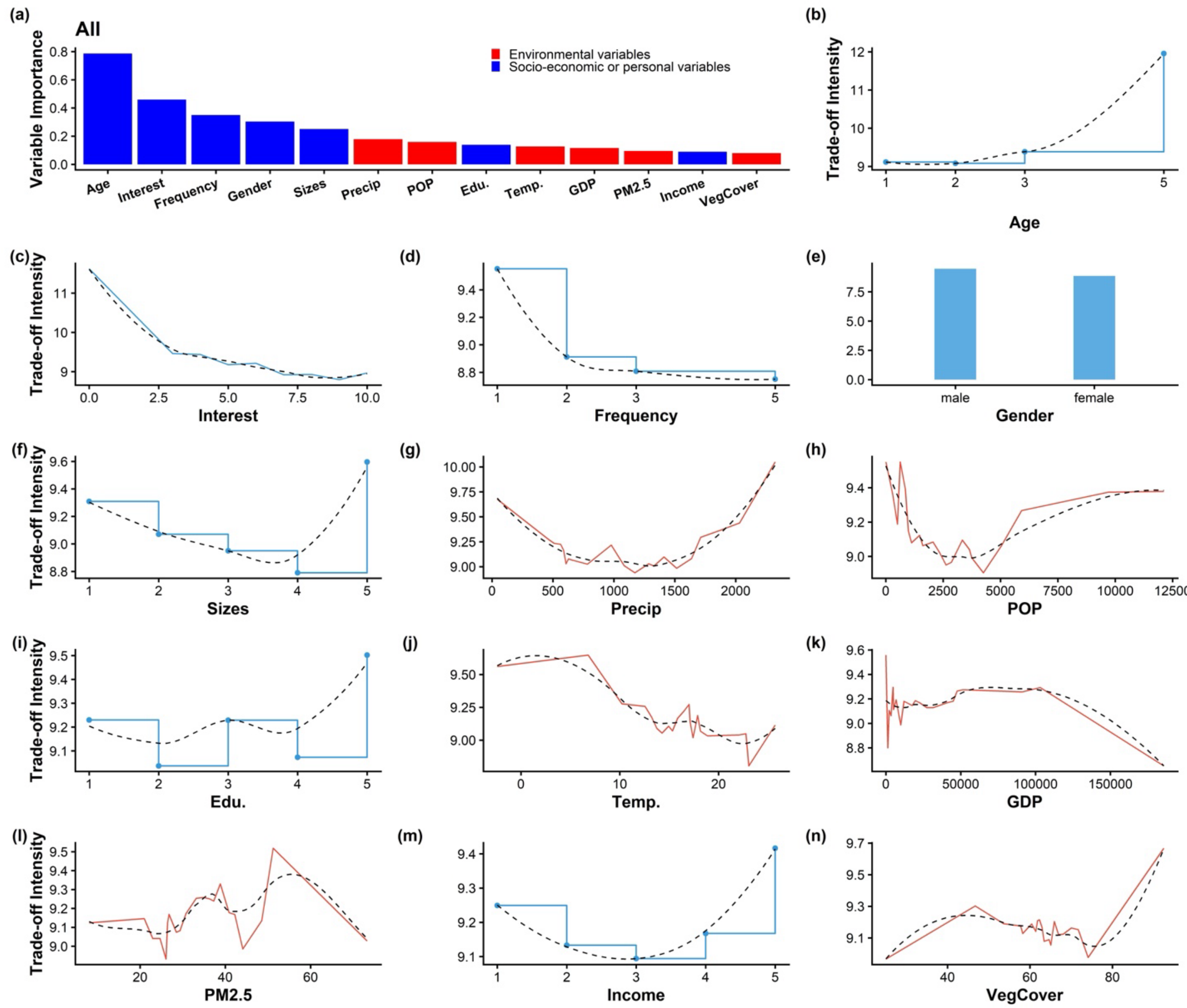

Fig. 4. Variable importance based on the XGBoost-SHAP analysis (a) and accumulated local effects plots of 13 socio-economic, personal, and environmental factors for predicting the trade-off intensity among ecosystem service demands (b-n). Black dotted curves are the nonparametric LOESS trendlines. Interest represents respondents' interest in environmental issues; Sizes represents the size of the most frequently visited park; Frequency is the monthly frequency of park visits; Edu. stands for education level; VegCover stands for vegetation coverage; POP stands for population density; Temp. stands for mean annual temperature; Precip. stands for mean annual precipitation.

## 4. Discussion

Our results suggested strong trade-offs between people's demands for Air Purification and Education, and for Air Purification and Food and Water Supply, in China's urban parks. This type of trade-off between regulating and provisioning services was also reported in previous supply-based studies (Lee, 2016; Yuan et al., 2023; Kirby et al., 2024). However, our demand-based approach highlights a crucial nuance: urban residents in China show an overwhelming preference for air quality improvement, even willing to forgo other potential benefits. This prioritization is evident in both pairwise trade-offs (Fig. 1a) and the emergence of a distinct "air purification-dominated" demand bundle (Fig. 2a). The high demand for the air purification service in China is probably related to China's history of the severe air pollution problem that caused millions of premature deaths and decrease in human happiness, especially before China launched the Clean Air Actions (Zheng et al., 2019; Yin et al., 2020; Geng et al., 2024). This historical context has cultivated a high level of public awareness and concern, translating into a strong societal demand for solutions, including nature-based approaches such as urban parks (Mexia et al., 2018).

Alongside air purification, our analysis also revealed a similarly strong preference for recreation services. (Fig. 1 and 2). Although recreation is a widely demanded urban ecosystem service worldwide (Berglihn and Gómez-Baggethun, 2021; Veerkamp et al., 2021; Patrycja et al., 2022), our study delineates a specific socio-demographic profile of the "recreation-dominated" group in China. This group typically comprises older individuals with higher income and education levels, who reside in densely populated, economically advanced cities with relatively better air quality (Fig. 3). This profile suggests that as basic needs for environmental quality are met and individuals gain more leisure time and financial resources, the demand for cultural ecosystem services such as recreation could become more pronounced (Fischer et al., 2018; Sun et al., 2019; Zhang and Zhou, 2018). The fact that these individuals live in cities with lower PM2.5 levels may also reduce barriers to outdoor activities, further reinforcing their demand for recreation (Subiza-Pérez et al., 2020). This dynamic implies that as China continues to develop and its environmental conditions improve, the demand for recreational services in urban parks is likely to grow, necessitating a shift in planning priorities.

Notably, the balanced demand bundle is the largest among the three identified groups, accounting for 58.69% of all respondents, which indicates that a majority of urban residents in China hold a comprehensive perception of the multiple ecosystem services provided by urban parks (Fig. 2a). The trade-off intensity of the balanced demand bundle is significantly lower than that of the other two bundles, forming a more stable demand structure for urban park services (although they still regard air purification as the relatively most important service but do not neglect other seven services) (Fig. 2b). Socio-demographically, it is dominated by younger females with strong environmental interests, compared with the air purification-dominated group (male-biased) and recreation-dominated group (elderly-biased), and geographically concentrated in cities with higher vegetation coverage and relatively lower GDP and population density (Fig. 3). These findings reflect the joint effects of high environmental literacy and sufficient green space supply on shaping comprehensive service perceptions (Chen et al., 2022; Jiang and Marggraf, 2025). Urban planners should prioritize building multi-functional urban parks in areas with such demand characteristics, integrating regulating (climate regulation), cultural (education), and supporting (habitat maintenance) services. Meanwhile, this group's high environmental awareness can be leveraged to promote ecological awareness, driving balanced service preferences in other groups.

In addition, our findings suggest that the intensity of trade-offs in ecosystem service demands is shaped by a complex interplay of personal attributes and broader environmental contexts (Fig. 4). The prominence of factors such as age and environmental interest, not only across all surveyed responses but also in air purification-dominated bundles (Fig. 4 and S6), underscores the socio-cultural dimension of ecosystem service valuation. Specifically, the finding that older individuals with lower environmental concern exhibit stronger trade-offs suggests a more focused demand for specific services, such as active recreation (Fig. 3 and 4). This is likely because these people may have more established routines and specific recreational or aesthetic preferences, leading them to prioritize a narrower set of services (Castro et al., 2014). Conversely, younger individuals with higher environmental engagement exhibited more balanced service preferences, reflecting possible generational shifts in environmental awareness and education (Martín-López et al., 2012). This also aligns with previous findings that knowledge about the natural environment is a key psychological determinant of people's perceptions of ES importance (Maleknia, 2024). The strong environmental interest of these people might provide them with greater motivation to learn more about the benefits people can obtain from nature and to engage more deeply with the natural environments of urban parks, which could lead to more balanced preferences among services .

Therefore, our findings supported the adoption of additional policies to expand environmental education and public engagement initiatives, thereby broadening public appreciation for the multifunctionality of urban parks.

Beyond individual characteristics, our results also demonstrate that city-level environmental factors establish a background context that modulates ESD trade-offs (Fig. 4, S5, and S7). The finding that trade-off intensity is lowest in cities with medium population densities is particularly insightful. Because the finding may imply the existence of a "sweet spot" where population density is high enough to foster high demand and appreciation for diverse park functions, but not so high as to create intensified conflicts between uses, leading to stronger trade-offs (Hansen et al., 2017; Zhan and Zheng, 2025). Similarly, the optimal ranges identified for mean annual temperature and precipitation likely reflect how regional climate shapes both the ES supply potential of parks and the types of outdoor activities that residents are accustomed to (Schirpke et al., 2024). The relatively low importance of vegetation coverage as a predictor was unexpected but highlights that the mere quantity of greenness may be less influential than its quality, accessibility, and configuration within the urban matrix (Shanahan et al., 2015; Borysiak and Stępniewska, 2022). Collectively, these results support place-based park governance: tailored design and outreach for older age groups, environmental education to broaden ES appreciation, and climate-sensitive zoning to sustain balanced multifunctionality can effectively reduce unintended ESD trade-offs and enhance equitable access to urban park benefits.

Last but not least, it should be acknowledged that our study has several limitations, which could open avenues for future research. First, the online survey sample is skewed toward younger, middle-aged, and more educated respondents from China's more developed eastern and central regions (Wu et al., 2026b) (Fig. S1). Consequently, the preferences of the elderly and residents in less-developed western provinces may be underrepresented. Second, our study provides a static snapshot of demand, whereas preferences are dynamic and context-dependent, varying by season, purpose of visit (e.g., purpose, time of day, season), and park type (e.g., small community park versus large forest park) (Zhao et al., 2024). Finally, our study design did not distinguish core urban from peri-urban residents within the self-reported ‘city of residence’. This is because residents commonly commute between urban and peri-urban areas, given the relatively well-developed intra-urban public transportation infrastructure in Chinese cities, thereby experiencing different ecosystem services across contexts (Sun and Cui, 2018). However, this design could introduce uncertainty into spatial comparison analyses and warrants a more careful distinction in fine-scale analyses. Future research should aim to address these limitations. For example, mixed-method approaches combining surveys with on-site interviews and behavioral observations could provide a more nuanced understanding of ESD. Targeted sampling strategies are needed to ensure the inclusion of underrepresented populations, such as the elderly and residents of remote or less developed regions, like Qinghai and Tibet. Furthermore, exploring the heterogeneity of demand across different types of urban green spaces (e.g., from small community parks to large-scale nature reserves, and park visitors in urban centers vs. rural areas) would yield more specific and actionable insights for planners and managers.

## 5. Conclusion

Our study provides the first national-scale results on the trade-offs among ecosystem service demands in urban parks across China. The findings suggest that three distinct groups of urban park visitors exist in China, each with unique service preferences: those who value recreation most, those who value cleaner air most, and those who prefer multiple services at once. The profile we delineated for the three types of people could inform the planning and design of urban parks in different neighborhoods to better match the demands of local residents. For instance, in areas with older populations or higher socio-economic status, planners should focus on providing diverse, high-quality recreational facilities, accessible pathways, and spaces for social gathering. Planners can also use our findings on the drivers of trade-off intensity as a diagnostic tool to anticipate public preferences based on local demographic and environmental data, thereby optimizing the allocation of limited resources to meet the community's most pressing needs. With this information, we could not only deepen our understanding of the nature of human-nature interactions but also create urban green infrastructure designs and management plans to sustainably enhance the quality of life for city residents around the world.